\documentclass[aps, prb, twocolumn, amssymb, amsmath, showpacs, superscriptaddress]{revtex4-1}
\usepackage{bm}
\usepackage{times}
\usepackage{graphicx}
\usepackage{color}
\usepackage{dcolumn}
\usepackage[colorlinks=true, letterpaper=true, pdfstartview=FitV, linkcolor=blue, citecolor=blue, urlcolor=blue]{hyperref}
\usepackage{appendix}
\usepackage[normalem]{ulem}

\newcommand{\vect}[1]{\boldsymbol{#1}}                 
\usepackage{helvet}
\usepackage{pifont}

\newcommand{\xmark}{\ding{55}}
\usepackage{multirow}
\usepackage{siunitx}
\usepackage{makecell}

\usepackage{colortbl}
\usepackage{xcolor}
\definecolor{lightgreen}{rgb}{0.88, 1.0, 0.88}

\begin{document}

\title{Third-order intrinsic anomalous Hall effect as a transport fingerprint of altermagnets}
\author{Longjun Xiang}
\email{xianglj@shanghaitech.edu.cn}
\affiliation{College of Physics and Optoelectronic Engineering, Shenzhen University, Shenzhen 518060, China}
\author{Hao Jin}
\affiliation{College of Physics and Optoelectronic Engineering, Shenzhen University, Shenzhen 518060, China}
\author{Jian Wang}
\email{jianwang@hku.hk}
\affiliation{College of Physics and Optoelectronic Engineering, Shenzhen University, Shenzhen 518060, China}
\affiliation{Quantum Science Center of Guangdong-Hongkong-Macao Greater Bay Area (Guangdong), Shenzhen 518045, China}
\affiliation{Department of Physics, The University of Hong Kong, Pokfulam Road, Hong Kong, China}


\begin{abstract}
The intrinsic anomalous Hall effect (IAHE) provides a powerful transport fingerprint of quantum magnets,
with its linear and second-order responses distinguishing ferromagnets and $\mathcal{P}\mathcal{T}$-symmetric antiferromagnets, respectively.
Altermagnets, as an emergent class of quantum magnets, have recently been shown to host
a third-order extrinsic anomalous Hall effect,
raising a question of whether an \textit{intrinsic} counterpart can serve as a diagnostic of altermagnetic order.
Based on spin-group symmetry analysis, we demonstrate that the third-order IAHE is generically allowed
in the ten spin Laue groups relevant to altermagnets when spin-orbit coupling (SOC) is taken into account.
By combining these symmetry constraints with the anomalous velocity induced by the second-order Berry curvature, 
we uncover a resonant third-order IAHE arising near the altermagnetic band crossings at generic momenta
in both the Lieb-lattice altermagnet and the experimentally realized altermagnet V$_2$Se$_2$O.
Notably, we identify the Berry curvature quadrupole,
encoded in the second-order Berry curvature and activated by finite SOC,
as the microscopic quantum geometric origin of this resonance.
Our results establish the third-order IAHE as an intrinsic quantum geometric transport fingerprint of altermagnets
and extend the hierarchy of IAHE across collinear quantum magnets.
\end{abstract}

\maketitle

\begin{figure}[t!]
\includegraphics[width=0.90\columnwidth]{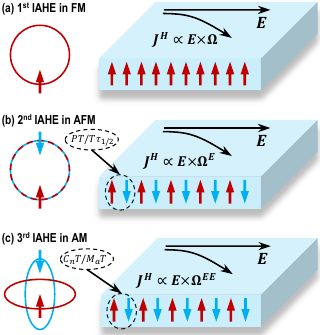}
\caption{
The hierarchy of IAHE across collinear quantum magnets.
(a) The ferromagnet (FM) with a spin-polarized Fermi-surface can be identified by the linear IAHE.
(b) The $\mathcal{P}\mathcal{T}$-symmetric antiferromagnet (AFM) with a spin degenerate Fermi surface
can be identified by the second-order IAHE.
(c) The altermagnet (AM) with an alternative spin-splitting Fermi surface can be identified by the third-order IAHE (This work).
Note that the opposite-spin sublattices in AFM are connected with $\mathcal{P}\mathcal{T}$ or $\mathcal{T}\tau_{1/2}$ symmetry,
while the opposite-spin sublattices in AM are connected with $C_n\mathcal{T}$ or $M_{a}\mathcal{T}$ symmetry~\cite{AM2}.
}
\label{FIG1}
\end{figure}

\noindent \textit{\textcolor{blue}{Introduction.}}---
The emergence of altermagnets opens a new chapter in quantum magnets~\cite{AM1, AM2, AM3, AM4, AM41, AM5, AM6, AM7, AM8, AM9},
where a variety of physical effects, such as Coulomb drag~\cite{Drag2025}, photocurrent~\cite{photocurrent},
and linear magnetoresistance~\cite{MR}, have been proposed to capture their fingerprints.
Historically, ferromagnetic materials with a spin-polarized Fermi surface can be
identified by the linear intrinsic anomalous Hall effect (IAHE)~\cite{Nagaosa2010, Xiao2010},
as shown in Fig.~\ref{FIG1}(a).
Conventional antiferromagnets with a spin-degenerate Fermi surface, on the other hand,
can be diagnosed using the second-order IAHE~\cite{GaoY2014PRL, BPT1, BPT2, XuSY2023, Wang2023, Han2024},
as shown in Fig.~\ref{FIG1}(b).
Recently, the third-order extrinsic anomalous Hall effect (EAHE)~\cite{KTLawPRB, Nanotech2021}
has been proposed as a probe for altermagnets~\cite{alter2025},
where both linear and second-order IAHEs are forbidden by symmetry.
However, the third-order IAHE~\cite{xiangThird, AmitThird, Neupert2025} in altermagnets,
particularly as the lead-order contribution,
remains largely unexplored~\cite{alter2024, alter2026}.

Within the semiclassical framework,
all the intrinsic anomalous Hall currents can be formally derived from the anomalous velocity~\cite{Xiao2010}.
As illustrated in Fig.~\ref{FIG1}, the anomalous velocity for linear IAHE is governed by the unperturbed Berry curvature (BC)~\cite{Xiao2010} $\vect{\Omega}$,
whereas those for the second- and third-order IAHEs originate from the first-order BC~\cite{GaoY2014PRL} $\vect{\Omega}^{\vect{E}}$
and the second-order BC~\cite{xiangThird} $\vect{\Omega}^{\vect{EE}}$, respectively.
The first-order BC is linked to the underlying quantum metric dipole (QMD)~\cite{BPT1, BPT2, XuSY2023, Wang2023, Han2024},
serving as the quantum geometric origin of the second-order IAHE.
In contrast, the underlying quantum geometric origin encoded in the second-order BC remains elusive.
Furthermore, while spin–orbit coupling (SOC) is essential for the emergence of both BC and QMD,
how it intertwines with the quantum geometry underlying the third-order IAHE remains largely unexplored,
particularly in altermagnets where SOC can be vanishingly small.

In this Letter, by using the spin-group symmetry analysis,
we demonstrate that the third-order IAHE is a spin-group-symmetry-breaking effect,
analogous to the linear IAHE~\cite{GaoPRX2025},
and that it can arise in the ten spin Laue groups~\cite{AM2} relevant to altermagnets
once SOC is included, even when SOC is weak.
By combining these symmetry constraints with the anomalous velocity associated with the second-order BC,
we develop a quantum-geometric decomposition of the third-order IAHE
and unveil its resonant enhancement near the altermagnetic band crossings at generic momenta
in both the Lieb-lattice altermagnet~\cite{Feng2026MOT, Lieb1, Lieb2} 
and the experimentally realized altermagnet V$_2$Se$_2$O~\cite{AM4, V2X2O, VXO1, VXO2, VXO3}.
Moreover, we identify the Berry curvature quadrupole (BCQ)~\cite{KTLawPRB, BCQ2, BCQ3, BCQ4},
encoded in the second-order BC and induced by a nonvanishing SOC,
as the quantum-geometric origin of this resonance.
For the altermagnetic band crossings at high-symmetry momenta,
we also find a resonant third-order IAHE governed by the acceleration quantum metric dipole (AQMD),
which is encoded in the second-order BC but requires strong SOC to be activated.
Our results establish the third-order IAHE as a transport fingerprint of altermagnets
and extend the hierarchy of IAHE across collinear quantum magnets, as summarized in Fig.~\ref{FIG1}.

\bigskip
\noindent{\textit{\textcolor{blue}{Quantum-geometric decomposition of third-order IAHE.}}}---The third-order
intrinsic anomalous Hall current density $\vect{j}$ in a DC electric field $\vect{E}$ can be written as $j_a = \sigma_{abcd}^{(0)} E_b E_c E_d$.
Here the Hall nature of this response requires $\sigma_{abcd}^{(0)} = - \sigma_{bacd}^{(0)}$,
which ensures a dissipationless transverse current~\cite{Nagaosa2010} with $j_aE_a=0$.
Besides, the intrinsic character~\cite{Xiao2010} of this response is reflected by its independence 
of the relaxation time $\tau$, namely $\sigma_{abcd}^{(0)} \propto \tau^0$.
Throughout this work, we adopt the Einstein summation convention for the
repeated Cartesian coordinates $a, b, c, d \in \{x, y, z\}$.

Within the semiclassical theory up to third order in the applied electric field $\vect{E}$,
the third-order IAHE can be expressed as
$j_a \equiv -e^2/\hbar \sum_n \int_k f_n (\vect{E} \times \vect{\Omega}^{\vect{E}\vect{E}}_n)_a$,
where $\Omega^{b;\vect{EE}}_n=\epsilon_{bcd}\partial_c T_n^{dij}E_iE_j$ is the second-order BC for the $n$th Bloch band.
The corresponding conductivity tensor reads~\cite{xiangThird}
\begin{align}
\sigma_{abcd}^{(0)} = \dfrac{e^2}{\hbar} \sum_n \int_k f_n' \left( v_n^a T_n^{bcd} - v_n^b T_n^{acd} \right),
\label{tau0}
\end{align}
where $\int_k = \int d^dk/(2\pi)^d$ with $d$ the spatial dimension,
$f_n'=\partial f_n/\partial \epsilon_n$,
with $f_n$ and $\epsilon_n$ being the equilibrium Fermi distribution function and the $n$th band energy, respectively,
and $v_n^a=\partial_a \epsilon_n$ with $\partial_a \equiv \partial/\partial k_a$ is the band velocity.
In addition, $T_n^{bcd}$ is the second-order Berry-connection polarizability tensor (BPT)~\cite{xiangThird, Jia2024},
which can be decomposed as (see End Matter)
\begin{align}
T_{n}^{bcd} 
&=
e^2 \sum_{m} \dfrac{5 \Omega^{bc}_{nm}\Delta^d_{nm} + 5 \Omega^{bd}_{nm}\Delta^c_{nm}}{- 4 \epsilon_{nm}^3}
\nonumber \\
&+
e^2 \sum_{m} \dfrac{4\Pi^{cdb}_{nm}+\Pi^{bcd}_{nm}+\Pi^{bdc}_{nm}}{-2\epsilon_{nm}^3}.
\label{Tbcd}
\end{align}
Here $\Omega^{bc}_{nm}=-2\text{Im}[r^b_{nm}r^c_{mn}]$ is the (local) BC,
where $r^b_{nm}$ with $n \neq m$ is the interband Berry connection;
$\Delta^d_{nm}=v_n^d-v_m^d$ is the group velocity difference of band $n$ and $m$;
$\Pi_{nm}^{cdb}=\text{Re}[w^{cd}_{nm}r^b_{mn}]$ with~\cite{SumRule} $w^{cd}_{nm}=\langle u_n|\partial^2_{cd}H|u_m\rangle$
is defined as the acceleration quantum metric
since $\partial^2_{cd}H$ ($H$, the Bloch Hamiltonian) is the acceleration operator.
Substituting Eq.~\eqref{Tbcd} into Eq.~(\ref{tau0}) gives the quantum-geometric decomposition of the third-order IAHE,
which arises from the underlying BCQ~\cite{KTLawPRB, BCQ2, BCQ3, BCQ4} but in the form of $v_n^av_n^d\Omega^{bc}_{nm}$
and the AQMD defined by $v_n^a\Pi^{bcd}_{nm}$.

Eq.~(\ref{tau0}), together with a quantum-geometric decomposition determined by the second-order BPT Eq.~\eqref{Tbcd},
is one of the main results of this work. Some remarks are in order.
First, the factor $f'_n$ in Eq.~\eqref{tau0} makes the third-order IAHE a Fermi-surface response,
so it occurs only in metals~\cite{BPT1}.
This is in stark contrast with the recently reported third-order intrinsic Hall current~\cite{alter2024},
which shows a Fermi-sea property.
both BCQ and AQMD are gauge-invariant,
making Eq.~\eqref{tau0} suitable for evaluating realistic materials
when combined with first-principles calculations.
Third, both BCQ and AQMD are time-reversal-odd ($\mathcal{T}$-odd),
so $\sigma_{abcd}^{(0)}$ contributed by them is also $\mathcal{T}$-odd.
As a consequence, a nonzero third-order IAHE requires $\mathcal{T}$-symmetry breaking.
Finally, we note that the BCQ can also induce a $\mathcal{T}$-odd third-order EAHE~\cite{KTLawPRB, footnoteTeven},
as represented by $\sigma_{abcd}^{(2)}$ with $\sigma_{abcd}^{(2)} \propto \tau^2$ (see End Matter).
However, dimensional analysis gives~\cite{footnote3} $\sigma^{(0)}_{abcd}/\sigma^{(2)}_{abcd} \sim 40$
for light doping in a moderately dirty metal.
Consequently, the third-order IAHE is expected to dominate over the third-order EAHE in this regime.
To conclude this section, we mention that Eq.~\eqref{tau0}
can further contain three-band processes associated with the three-state quantum metric~\cite{MehraeenPRL} dipole (TQMD) (see End Matter),
but whose contribution is negligibly small. 

\bigskip
\noindent{\textit{\textcolor{blue}{Third-order IAHE being a spin-group-symmetry-breaking effect.}}}---Based on
spin-group symmetry analysis, we show that the third-order IAHE requires both $\mathcal{T}$-symmetry breaking and spin-group-symmetry breaking,
and is generically allowed in the ten spin Laue groups~\cite{AM2} relevant to altermagnets in the presence of SOC.
To make the symmetry constraints transparent,
we transform the antisymmetric rank-4 tensor $\sigma^{(0)}_{abcd}$
into a rank-3 pseudotensor $\kappa_{abc}$~\cite{spinless}.
Treating SOC as a perturbation,
we parametrize it by the spin-orbit tensor~\cite{GaoPRX2025} $\ell_{i}^\alpha$,
where $i \in \{x, y, z\}$ and $\alpha=1, 2, 3$ denote the Cartesian coordinates in real space
and a color index in spin space, respectively,
and expand $\kappa_{abc}$ as~\cite{GaoPRX2025, Feng2026MOT},
\begin{align}
\kappa_{abc}
=
\kappa_{abc}^{(0)}
+
\sum_{i}^{\alpha}
\kappa_{abc;i}^{(1);\alpha} \ell_i^\alpha
+
\sum_{ij}^{\alpha\beta}
\kappa_{abc;ij}^{(2);\alpha\beta} \ell_i^\alpha \ell_j^\beta
+
\cdots.
\label{socexp}
\end{align}
The symmetry constraints on $\kappa_{abc}$ can then be obtained systematically
by applying the spin point-group operations order by order to the expansion coefficients.

In general, a spin group can be decomposed into the spin-only subgroup $r_s$ and
a nontrivial spin subgroup $R_s$, whose elements are denoted by $[u||R]$~\cite{AM2},
where $u$ and $R$ act in spin and real space, respectively.
For altermagnets with a N\'eel vector $\vect{N}$ along the $z$-direction,
the spin-only group can be generated by~\cite{AM2} $[\bar{C}_{2}^{a}||E]$ ($a=x, y$) and $[C_{\varphi}^z||E]$,
where $E$ is the identity operation, $\bar{C}_n^a$ is the $n$-fold rotation along the $a$-direction in spin space,
followed by an inversion in the same space,
and $C_\varphi^z$ is a continuous rotation along the $z$-direction in spin space.
Under $[\bar{C}_2^a||E]$, which effectively plays the role of $\mathcal{T}$-symmetry~\cite{AM2} on $\kappa_{abc}$,
we obtain $\kappa_{abc}^{(0)}=-\kappa_{abc}^{(0)}$ so that
the third-order IAHE vanishes identically in the absence of SOC.
Therefore, the third-order IAHE should be regarded not merely as a
$\mathcal{T}$-breaking effect, but as a spin-group-symmetry-breaking effect,
analogous to the linear IAHE~\cite{GaoPRX2025}.

Moreover, by applying the two spin-only operations on the remaining terms of Eq.~\eqref{socexp},
up to second order in SOC, we find~\cite{GaoPRX2025, Feng2026MOT}
$\kappa_{abc} 
=
\kappa_{abc;i}^{(1);3}\ell_{i}^3 + \kappa_{abc;ij}^{(2);12}(\ell_{i}^1\ell_{j}^2-\ell_{i}^2\ell_{j}^1)
$,
where $\ell_i^3$ along the N\'eel vector $\vect{N}$ is responsible for the spin-conserving SOC
whereas $\ell_i^{1}$ and $\ell_i^{2}$ correspond to the spin-flip SOC.
Beyond the spin-only group, the nontrivial spin group imposes additional constraints on
$\kappa_{abc;i}^{(1);\alpha}$ and $\kappa_{abc;ij}^{(2);\alpha\beta}$ through~\cite{GaoPRX2025, Feng2026MOT}
$
\kappa_{abc;i}^{(1);\alpha} = R_{aa'}R_{bb'}R_{cc'} R_{ii'} u_{\alpha\alpha'} \kappa_{a'b'c;i'}^{(1);\alpha'}
$
and
$
\kappa_{abc;ij}^{(2);\alpha\beta} = \eta_T |R| R_{aa'}R_{bb'}R_{cc'} R_{ii'} R_{jj'} 
u_{\alpha\alpha'} u_{\beta\beta'}
\kappa_{a'b'c';i'j'}^{(2);\alpha'\beta'}
$, respectively,
where $\eta_T =\pm 1 $ are for $R$ and $R\mathcal{T}$,
respectively, and $|R|$ is the determinant of $R$.
The results for $\kappa_{abc;i}^{(1);3}$ and $\kappa_{abc;ij}^{(2);12}$
with $\vect{j} \perp \vect{N}$ and $\vect{E}\perp \vect{N}$ are summarized in Table~\ref{TableI}.

We notice that the third-order IAHE can be allowed by all the ten spin Laue groups~\cite{footnote1}
relevant to altermagnets when SOC is considered.
Importantly, this response becomes the leading-order IAHE
in the spin Laue groups $^24/m$, $^24/^1m^2m^1m$, $^26/^2m$, $^26/^2m^2m^1m$, and $^1m^1\bar{3}^2m$,
which correspond to many experimentally confirmed altermagnets,
including V$_2$Se$_2$O~\cite{V2X2O, VXO1, VXO2, VXO3} and RuO$_2$~\cite{RuO21, RuO22, RuO23, RuO24} with $^24/^1m^2m^1m$,
CrSb~\cite{CrSb1, CrSb2, CrSb3, CrSb4} and MnTe~\cite{MnTe1, MnTe2, MnTe3} with $^26/^2m^2m^1m$.
To close this section, we remark that the perturbative spin-group tensors listed in Table~\ref{TableI}
are consistent with the nonperturbative magnetic point group analysis~\cite{xiangThird}
and the above spin-group symmetry analysis can be
applied to $\sigma_{abcd}^{(2)}$ for the third-order EAHE.

\begin{table}[t!]
\centering
\caption{The allowed elements of $\kappa^{(1);3}_{abc;i}$ and $\kappa^{(2);12}_{abc;ij}$ for the ten spin Laue groups,
particularly with the N\'eel vector $\vect{N}$ along the $z$-direction.
Note that $\kappa^{(1);3}_{abc;i}=\kappa^{(1);3}_{acb;i}$ and $\kappa^{(2);12}_{abc;ij}=\kappa^{(2);12}_{acb;ij}$.
Here, \xmark~means that no component is allowed by symmetry.
For simplicity, we only listed the elements with $\vect{j} \perp \vect{N}$ and $\vect{E} \perp \vect{N}$.
\label{TableI}
}
\renewcommand{\arraystretch}{1.5}
\begin{tabular}{lll}
\hline\hline
\ \ Spin Laue group \ \ \
& \ \ \ $\kappa^{(1);3}_{abc;i}$ \ \ \ 
& \ \ \ $\kappa^{(2);12}_{abc;ij}$ \\
\hline
\ \ $^2m^2m^1m$ \ \ \ 
& \ \ \ $(zxx;z), (zyy;z)$ \ \ \
& \ \ \ $(zxx;xy), (zyy;xy)$ \\

\ \ $^24/^1m$ \ \ \
& \ \ \ $(zxx=-zyy;z)$ \ \ \
& \ \ \ $(zxx=-zyy;xy)$ \\

\ \           \ \ \
& \ \ \ $(zxy;z)$ \ \ \
& \ \ \ $(zxy;xy)$ \\

\ \ $^24/^1m^2m^1m$ \ \ \
& \ \ \ $(zxx=-zyy;z)$ \ \ \
& \ \ \ $(zxx=-zyy;xy)$ \\

\ \ $^14/^1m^2m^2m$ \ \ \
& \ \ \ $(zxx=zyy;z)$ \ \ \
& \ \ \ $(zxx=zyy;xy)$ \ \ \ \\

\ \ $^16/^1m^2m^2m$ \ \ \
& \ \ \ $(zxx=zyy;z)$ \ \ \
& \ \ \ $(zxx=zyy;xy)$ \\

\ \ $^22/^2m$ \ \ \
& \ \ \ $(zxx;z), (zyy;z)$ \ \ \
& \ \ \ $(zxx;xy), (zyy; xy)$ \\

\ \ $^1\bar{3}^2m$ \ \ \
& \ \ \ $(zxx=zyy;z)$ \ \ \
& \ \ \ $(zxx=zyy;xy)$ \\

\ \ $^26/^2m$ \ \ \ 
& \ \ \  \xmark  \ \ \
& \ \ \ \xmark \\

\ \ $^26/^2m^2m^1m$ \ \ \
& \ \ \ \xmark \ \ \
& \ \ \ \xmark \\

\ \ $^1m^1\bar{3}^2m$ \ \ \
& \ \ \ $(zxy;z)$
& \ \ \ $(zxy;xy)$ \\
\hline\hline
\end{tabular}
\end{table}

\begin{figure}[t!]
\includegraphics[width=0.90\columnwidth]{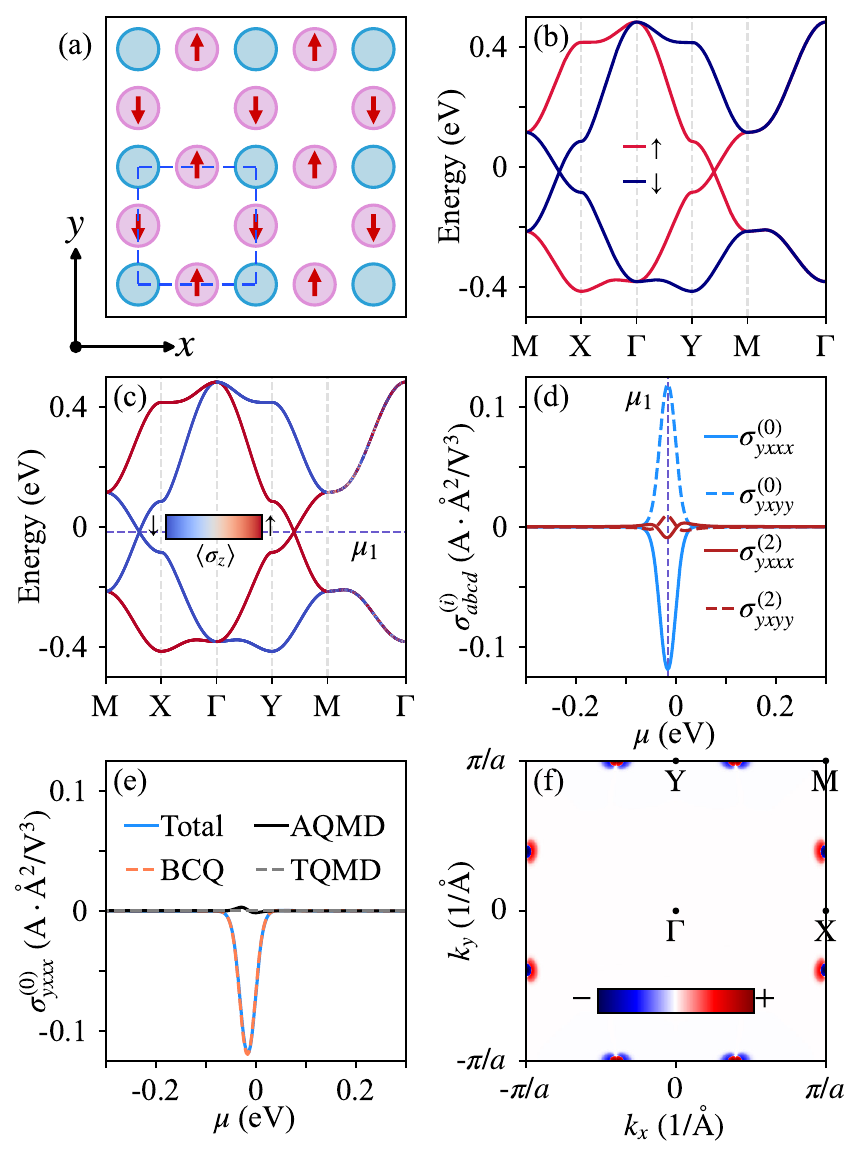}
\caption{
Third-order IAHE in 2D Lieb-lattice altermagnet.
(a) The 2D Lieb-lattice. Here the blue dashed square labels the unit cell.
(b) The spin-polarized band dispersions of Eq.~(\ref{LiebH}) with $\lambda_z=0$.
(c) The band dispersions of Eq.~(\ref{LiebH}) with $\lambda_z=0.01t_1$.
Here the colorbar denotes the expectation value of $\sigma_z$.
(d) The dependence of $\sigma^{(0/2)}_{yxxx}$ and $\sigma^{(0/2)}_{yxyy}$ on the chemical potential $\mu$.
(e) The quantum-geometric decompositions of $\sigma^{(0)}_{yxxx}$.
(f) The normalized $\vect{k}$-resolved integrand (contributed by BCQ) for the resonant peak $\mu_1$ of $\sigma_{yxxx}^{(0)}$.
Parameters: $4t_1=-0.4 \mathrm{eV}$, $2t_2=+0.025 \mathrm{eV}$, $2t_d=+0.125 \mathrm{eV}$, $J=-0.165 \mathrm{eV}$,
$T=100 \mathrm{K}$, and $\tau=10^{-14} \mathrm{s}$.
}
\label{FIG2}
\end{figure}

\bigskip
\noindent{\textit{\textcolor{blue}{Third-order IAHE in Lieb-lattice altermagnet.}}}---Guided by the spin-group symmetry analysis,
we first explore the third-order IAHE in the 2D Lieb-lattice altermagnet with a spin Laue group $^24/^1m^2m^1m$ in the absence of SOC.
Its unit cell (denoted by a blue dashed square) is shown in Fig.~\ref{FIG2}(a), 
which contains two opposite-spin magnetic sites at $(0, 1/2)$ and $(1/2, 0)$, related by a fourfold rotation,
and a nonmagnetic site at $(0, 0)$.
By projecting out the nonmagnetic site and considering the nearest-neighbor hopping $t_1$ and
the anisotropic next-nearest-neighbor (NNN) hopping $t_2^{\pm}=t_2 \pm t_d$
between magnetic sublattices, the Bloch Hamiltonian reads~\cite{Feng2026MOT, Lieb1, Lieb2}
\begin{align}
H &= t_1 h_1 \tau_x + t_2h_2 \tau_0 + t_d h_3 \tau_z+J\tau_z\sigma_z + \lambda_zh_4\tau_y\sigma_z,
\label{LiebH}
\end{align}
where $\tau_i$ and $\sigma_i$ with $i\in\{x, y, z\}$ are Pauli matrices acting on the sublattice and spin spaces, respectively,
$\tau_0$ is a two by two identity matrix,
$h_1=4\cos (k_x/2) \cos (k_y/2)$, $h_2=2(\cos k_x+ \cos k_y)$, $h_3=2(\cos k_x - \cos k_y)$,
and $h_4=4\sin (k_x/2) \sin (k_y/2)$.
In addition, $J$ and $\lambda_z$ represent
the staggered exchange field from the N\'eel order and the spin-conserving SOC, respectively.
For brevity, we set N\'eel vector $\vect{N}$ along $z$-direction below.

The spin-polarized band dispersions for this model with $\lambda_z=0$ are shown in Fig.~\ref{FIG2}(b),
where the altermagnetic spin splitting (along $\mathrm{M}-\mathrm{X}-\Gamma-\mathrm{Y}-\mathrm{M}$)
arising from the anisotropic NNN hopping is observed.
This splitting is anisotropic in momentum space due to the presence of the nodal line along $\Gamma-\mathrm{M}$,
which is protected by the spin-group symmetry~\cite{Feng2026MOT} $[C_2||\mathcal{M}_{xy}]$.
After switching on the spin-conserving SOC with a small amplitude $\lambda_z=0.01t_1$,
we find that the band dispersion of Eq.~\eqref{LiebH} in Fig.~\ref{FIG2}(c) is almost unchanged,
as can be seen from the intact spin polarization of $\sigma_z$ [see the inset colorbar in Fig.~\ref{FIG2}(c)].

In the presence of SOC (even if weak), however, we note that the Lieb-lattice altermagnet with $^24/^1m^2m^1m$
permits $\kappa_{zxx}=-\kappa_{zyy}$ or $-\sigma_{yxxx}^{(0)}=\sigma_{yxyy}^{(0)}$
for the third-order IAHE (see Table~\ref{TableI}).
The dependence of $\sigma_{yxxx}^{(0)}$ and $\sigma_{yxyy}^{(0)}$ on the chemical potential $\mu$ at a temperature of $100 \mathrm{K}$
is shown in Fig.~\ref{FIG2}(d), where $\sigma_{yxxx}^{(2)}$ and $\sigma_{yxyy}^{(2)}$ are plotted for comparison.
Consistent with the symmetry analysis,
both the intrinsic and extrinsic components satisfy $\sigma_{yxxx}^{(0/2)}=-\sigma_{yxyy}^{(0/2)}$.
In addition, we find that the third-order EAHE with $\tau=10^{-14} \mathrm{s}$
is much smaller than its intrinsic counterpart, consistent with the dimensional analysis~\cite{footnote3}.
Importantly, we find that both $\sigma_{yxxx}^{(0)}$ and $\sigma_{yxyy}^{(0)}$
feature a resonant enhancement when $\mu$ is located near the altermagnetic band crossings
at generic momenta along $\mathrm{X}-\mathrm{M}$ and $\mathrm{Y}-\mathrm{M}$,
as marked by the purple dashed line $\mu_1$ in Figs.~\ref{FIG2}(c) and \ref{FIG2}(d).
We remark that the two altermagnetic band crossings
will be gapped out (topological band inversion) by a spin-conserving SOC (such as $\lambda_z=0.1t$~\cite{Feng2026MOT})
to drive the system into a mirror Chern insulator~\cite{Lieb2}.
This explains why they can produce a pronounced quantum-geometric enhancement even in the weak-SOC limit.
Furthermore, using the quantum-geometric decomposition of the third-order IAHE,
we show that the resonant peak near $\mu_1$ is dominated by BCQ, as shown in Fig.~\ref{FIG2}(e),
where the contributions from both AQMD and TQMD are negligible.
Finally, we show the normalized $\vect{k}$-resolved integrand contributed by BCQ for $\sigma_{yxxx}^{(0)}$ at $\mu_1$
in Fig.~\ref{FIG2}(f), confirming that the resonant enhancement indeed
arises from the topological altermagnetic band crossings at generic momenta.

In addition to the spin-conserving SOC, we can introduce a spin-flip SOC~\cite{Feng2026MOT}
$H_1=\lambda_{\parallel}\left(h_5 \tau_y\sigma_x+h_6 \tau_y\sigma_y \right)$ in Eq.~\eqref{LiebH},
where $h_5=4 \sin (k_x/2) \cos (k_y/2)$ and $h_6=4 \cos (k_x/2) \sin(k_y/2)$.
This SOC lifts the nodal line along $\Gamma-\mathrm{M}$ and can induce four additional resonant peaks in $\sigma_{yxxx}^{(0)}$,
which are dominated by AQMD at high-symmetry momenta
but need a large SOC, see End Matter.

\begin{figure}[t!]
\includegraphics[width=0.90\columnwidth]{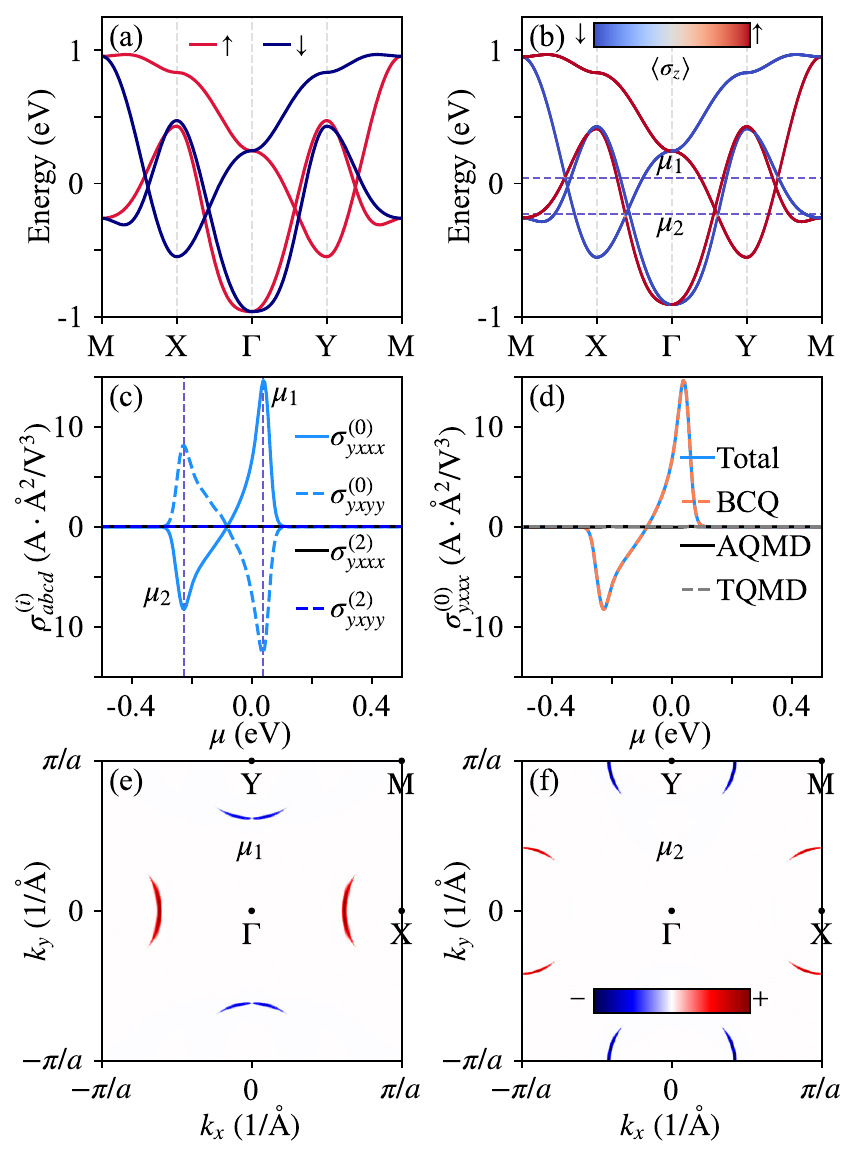}
\caption{
Third-order IAHE in van der Waals altermagnet V$_2$Se$_2$O.
(a) and (b): The band dispersions of V$_2$Se$_2$O$_2$ without and with SOC, respectively.
(c) The dependence of $\sigma^{(0/2)}_{yxxx}$ and $\sigma^{(0/2)}_{yxyy}$ on the chemical potential $\mu$.
To evaluate $\sigma^{(2)}_{abcd}$, we choose $\tau=10^{-14}\mathrm{s}$.
(d) The quantum-geometric decomposition of $\sigma^{(0)}_{yxxx}$.
(e) and (f): The normalized $\vect{k}$-resolved integrand (contributed by BCQ)
for the $\mu_1$ and $\mu_2$ resonant peaks of $\sigma^{(0)}_{yxxx}$.
}
\label{FIG3}
\end{figure}

\bigskip
\noindent{\textit{\textcolor{blue}{Third-order IAHE in van der Waals altermagnet V$_2$Se$_2$O}.}}---Beyond model studies,
we demonstrate the third-order IAHE in the experimentally confirmed van der Waals altermagnet V$_2$Se$_2$O~\cite{VXO1, VXO2, VXO3}.
The low-energy electronic structure of V$_2$Se$_2$O
can be described by a first-principles four-band intralayer tight-binding model
constructed from the dominant V-$d$ orbitals near the Fermi level~\cite{V2X2O}
\begin{equation}
H=
\begin{pmatrix}
H_{xz,\uparrow} & 0 & 0 & h_{03} \\
0 & H_{xy,\uparrow} & h_{12} & 0 \\
0 & h_{21} & H_{yz,\downarrow} & 0 \\
h_{30} & 0 & 0 & H_{xy,\downarrow}
\end{pmatrix},
\label{H0}
\end{equation}
where the diagonal terms give rise to the altermagnetic spin splitting
while the off-diagonal terms represent the vanishingly small SOC in V$_2$Se$_2$O.
The matrix elements of Eq.~\eqref{H0} are provided in the Supplementary Material~\cite{sup}.
The spin-polarized band dispersions [Fig.~\ref{FIG3}(a)] of Eq.~\eqref{H0}
exhibit a large altermagnetic spin splitting along $\mathrm{M}-\mathrm{X}-\Gamma-\mathrm{Y}-\mathrm{M}$,
which is also anisotropic due to the nodal line along $\Gamma-\mathrm{M}$ enforced by spin-group symmetry (not shown).
Upon including SOC, the band dispersions [Fig.~\ref{FIG3}(b)] remain almost unchanged,
as confirmed by the intact spin polarization of $\sigma_z$ [see the inset colorbar in Fig.~\ref{FIG3}(b)].

Notably, V$_2$Se$_2$O hosts the same spin Laue group as the Lieb-lattice altermagnet in the absence of SOC,
so that $\sigma^{(0/2)}_{yxxx}$ and $\sigma^{(0/2)}_{yxyy}$ are symmetry allowed when SOC is included.
Their dependence on $\mu$ is shown in Fig.~\ref{FIG3}(c),
where the $\sigma^{(2)}_{yxxx}$ and $\sigma^{(2)}_{yxyy}$, with $\tau = 10^{-14}\mathrm{s}$, remain negligible
relative to their intrinsic counterparts, consistent with the dimensional analysis~\cite{footnote3}.
Importantly, both $\sigma_{yxxx}^{(0)}$ and $\sigma_{yxyy}^{(0)}$ exhibit resonant enhancement
when $\mu$ is tuned to $\mu_1$ and $\mu_2$ [vertical purple dashed lines in Fig.~\ref{FIG3}(c)],
which lie near the altermagnetic band crossings [horizontal purple dashed lines in Fig.~\ref{FIG3}(b)]
at generic momenta along $\mathrm{X}$–$\mathrm{M}$ and $\mathrm{Y}$–$\mathrm{M}$.
Instead of being gapped by SOC,
these altermagnetic band crossings are strongly hybridized even for the relatively weak SOC in V$_2$Se$_2$O,
and hence underpin the observed enhancement.
Taking $\sigma^{(0)}_{yxxx}$ as a representative example,
we find that its quantum geometric enhancement is overwhelmingly dominated by BCQ, as shown in Fig.~\ref{FIG3}(d).
Besides, we confirm that the enhancement near $\mu_1$ and $\mu_2$
arises from the altermagnetic band crossings at generic momenta,
by plotting the normalized $\vect{k}$-resolved integrand (from BCQ) of $\sigma^{(0)}_{yxxx}$
in Figs.~\ref{FIG3}(e) and \ref{FIG3}(f), respectively.

\bigskip
\noindent \textit{\textcolor{blue}{Summary.}}---In summary,
we have established the third-order IAHE as a transport fingerprint of altermagnets,
thereby extending the hierarchy of IAHE across collinear quantum magnets.
Our symmetry analysis reveals that this response originates from spin-group symmetry breaking induced by SOC,
and is generically allowed in all ten spin Laue groups relevant to altermagnets.
Remarkably, employing the quantum-geometric decomposition of third-order IAHE,
we show that the third-order IAHE can exhibit a BCQ-dominated
resonant enhancement near altermagnetic band crossings at generic momenta.
Given its intrinsic nature, our results offer a robust and experimentally
accessible probe of altermagnetic order in the moderately dirty regime~\cite{Nagaosa2010}.
Particularly, we find that the resonant peak in V$_2$Se$_2$O,
with $\sigma_{yxxx}^{(0)} \sim 10(\mathrm{A}\mathrm{\mathring{A}}^2/\mathrm{V}^3)$ at a temperature of $100\mathrm{K}$,
gives a Hall voltage of $10 (\mu\mathrm{V})$,
as estimated by considering an applied electric field of $\sim 10^{5} (\mathrm{V/m})$,
a sample resistance of $\sim 10^{3} (\Omega)$, and a Hall-bar lateral size of $\sim 100 (\mu\mathrm{m})$,
and making it experimentally detectable.

\bigskip
\section*{Acknowledgements}
We thank the National Natural Science Foundation of China (Grants No. 12404059 and No. 12034014).


\newpage
\section{End Matter}

\subsection{Quantum-geometric decomposition of the second-order BPT}
\label{twothree}
In an DC electric field, the second-order BPT is given by~\cite{xiangThird, Jia2024}
\begin{align}
\bar{T}_n^{bcd}
&=
e^2\text{Re} \sum_{m}
\frac{r_{mn}^{d}}{\epsilon_{mn}}
i\mathcal{D}_{nm}^{b}
\left(
\frac{r_{nm}^{c}}{\epsilon_{nm}}
\right)
\nonumber \\
&- 
e^2\text{Re} \sum_{m}
\dfrac{2r^b_{nm}}{\epsilon_{nm}}
i\mathcal{D}_{mn}^{d}
\left(
\dfrac{r_{mn}^{c}}{\epsilon_{mn}}
\right)
\nonumber \\
&+
e^2\sum_{m l}
\dfrac{2g^{bdc}_{nml}+g^{dbc}_{nml}}{\epsilon_{nl}\epsilon_{nm}}.
\label{Tbcdapp}
\end{align}
Here $\epsilon_{mn}=\epsilon_m-\epsilon_n$,
$r^d_{mn}$ is the interband ($m \neq n$) Berry connection,
as defined by $r^d_{mn}=\langle u_m|i\partial_d|u_n\rangle$ with $|u_m\rangle$ the cell-periodic Bloch state,
$\mathcal{D}_{nm}^b=\partial_b - i(\mathcal{A}_{nn}^b-\mathcal{A}_{mm}^b)$ is the $U(1)$ covariant derivative,
where $\mathcal{A}_{nn}^b=\langle u_n|i\partial_b|u_n\rangle$ is the intraband Berry connection,
and $g_{nml}^{bdc}=\text{Re}[r^b_{nm}r^d_{ml}r^c_{ln}]$ is the three-state quantum metric~\cite{MehraeenPRL}.

The last line of Eq.~(\ref{Tbcdapp}) is contributed by the three-band processes
while its first two lines are contributed by both the two-band and three-band processes.
Using the chain rule
\begin{align}
i\mathcal{D}_{nm}^b \left( \dfrac{r^c_{nm}}{\epsilon_{nm}} \right)
=
\dfrac{i\mathcal{D}_{nm}^b {r}^c_{nm}}{\epsilon_{nm}}
-
\dfrac{i r^c_{nm} \Delta^b_{nm} }{\epsilon_{nm}^2},
\end{align}
where $\Delta^b_{nm}=\partial_b \epsilon_{nm}$, and the sum rule~\cite{SumRule}
\begin{align}
i\mathcal{D}_{nm}^b {r}^c_{nm}
&=
\dfrac{-i(\Delta^b_{nm}r^c_{nm}+r^b_{nm}\Delta^c_{nm})+w^{bc}_{nm}}{\epsilon_{nm}}
\nonumber \\
&+
\sum_{l}
\dfrac{\epsilon_{nl}r^c_{nl}r^b_{lm} - r^b_{nl}r^c_{lm}\epsilon_{lm}}{\epsilon_{nm}},
\end{align}
where $w^{bc}_{nm}=\langle u_n|\partial^2_{bc}H|u_m\rangle$ with $H$ being the Bloch Hamiltonian,
the first line (denoted by $L_1$) of Eq.~(\ref{Tbcdapp}) can be simplified as:
\begin{align}
L_1 &= e^2 \sum_{m} \dfrac{2 \Omega^{cd}_{nm} \Delta^b_{nm} + \Omega^{bd}_{nm} \Delta^c_{nm} - 2\Pi_{nm}^{bcd}}{2\epsilon_{nm}^3}
\nonumber \\
&+
e^2 \sum_{ml} 
\dfrac{\epsilon_{ml}g^{bcd}_{nml}-\epsilon_{nm}g^{cbd}_{nml}}{\epsilon_{nl}^3},
\label{L1}
\end{align}
where $\Omega_{nm}^{cd}=-2\text{Im}[r^c_{nm}r^d_{mn}]$ is the (local) BC
while $\Pi_{nm}^{bcd}=\text{Re}[v^{bc}_{nm}r^d_{mn}]$ is defined as the (local) acceleration quantum metric.
Note that $\Omega_{nm}^{cd}=-\Omega_{mn}^{cd}=-\Omega_{nm}^{dc}=\Omega^{dc}_{mn}$
and $\Pi^{bcd}_{nm}=\Pi^{bcd}_{mn}=\Pi^{cbd}_{nm}=\Pi^{cbd}_{mn}$.
In addition, the three-band term is obtained by interchanging $l$ and $m$.
Similarly, the second line (denoted by $L_2$) of Eq.~(\ref{Tbcdapp}) can be simplified as:
\begin{align}
L_2 &= e^2 \sum_{m} \dfrac{2\Omega^{bc}_{nm} \Delta^d_{nm} + \Omega^{bd}_{nm} \Delta^c_{nm} + 2 \Pi_{nm}^{cdb}}{-\epsilon_{nm}^3}
\nonumber \\
&-
e^2 \sum_{ml} \dfrac{2(\epsilon_{ml}g^{bcd}_{nml}+\epsilon_{nl}g^{bdc}_{nml})}{\epsilon_{nm}^3}.
\label{L2}
\end{align}
Inserting Eqs.~(\ref{L1}) and \eqref{L2} into Eq.~\eqref{Tbcdapp}, we obtain
\begin{align}
\bar{T}_{n}^{bcd}
&=
e^2 \sum_{m} \dfrac{2 \Omega^{cd}_{nm} \Delta^b_{nm} - \Omega^{bd}_{nm} \Delta^c_{nm} - 4 \Omega^{bc}_{nm} \Delta^d_{nm}}{2\epsilon_{nm}^3}
\nonumber \\
&-
e^2 \sum_{m} \dfrac{2\Pi_{nm}^{cdb}+\Pi_{nm}^{bcd}}{\epsilon_{nm}^3}
+
e^2 \sum_{m l}
\dfrac{2g^{bdc}_{nml}+g^{dbc}_{nml}}{\epsilon_{nl}\epsilon_{nm}}
\nonumber \\
&-
e^2 \sum_{ml}
\dfrac{\epsilon_{ml}(g^{dcb}_{nml}+2g^{bcd}_{nml})}{\epsilon_{nm}^3}
\nonumber \\
&-
e^2 \sum_{ml}
\dfrac{\epsilon_{nl}(g^{dbc}_{nml}+2g^{bdc}_{nml})}{\epsilon_{nm}^3},
\end{align}
where we have used that $g^{bcd}_{nml}=\text{Re}[r^b_{nm}r^c_{ml}r^d_{ln}]=\text{Re}[r^b_{mn}r^c_{lm}r^d_{nl}]=g^{dcb}_{nlm}$.
Finally, by defining $T_n^{bcd}=(\bar{T}_n^{bcd}+\bar{T}^{bdc}_n)/2$ to symmetrize the indices $c$ and $d$, we obtain
\begin{align}
T_{n}^{bcd} 
&=
e^2 \sum_{m} \dfrac{5 \Omega^{bc}_{nm}\Delta^d_{nm} + 5 \Omega^{bd}_{nm}\Delta^c_{nm}}{- 4 \epsilon_{nm}^3}
\nonumber \\
&+
e^2 \sum_{m} \dfrac{4\Pi^{cdb}_{nm}+\Pi^{bcd}_{nm}+\Pi^{bdc}_{nm}}{-2\epsilon_{nm}^3}
\nonumber \\
&+
e^2 \sum_{ml} 
\left(
\alpha^{n}_{ml}g^{bcd}_{nml}
+
\beta^{n}_{ml}g^{cdb}_{nml}
+
\gamma^{n}_{ml}g^{dbc}_{nml}
\right),
\label{Tabcapp}
\end{align}
where
\begin{align}
\alpha_{ml}^n 
&=
\dfrac{1}{2}
\left(
\dfrac{2}{\epsilon_{nl}\epsilon_{nm}} + \dfrac{\epsilon_{ml}}{\epsilon_{nl}^3} - \dfrac{2\epsilon_{ml}}{\epsilon_{nm}^3}
-
\dfrac{2\epsilon_{nl}}{\epsilon_{nm}^3}
\right)
\nonumber \\
&=
\dfrac{(\epsilon_{nl}^2-\epsilon_{nm}^2)(4\epsilon_{nl}^2-\epsilon_{nm}^2-2\epsilon_{nl}\epsilon_{nm})}{-2\epsilon_{nl}^3\epsilon_{nm}^3},
\\
\beta_{ml}^n 
&=
\dfrac{1}{2}
\left(
\dfrac{2}{\epsilon_{nl}\epsilon_{nm}} - \dfrac{\epsilon_{ml}}{\epsilon_{nm}^3} + \dfrac{2\epsilon_{ml}}{\epsilon_{nl}^3}
-
\dfrac{2\epsilon_{nm}}{\epsilon_{nl}^3}
\right)
\nonumber \\
&=
\dfrac{(\epsilon_{nl}^2-\epsilon_{nm}^2)(\epsilon_{nl}^2+2\epsilon_{nl}\epsilon_{nm}-4\epsilon_{nm}^2)}{-2\epsilon_{nl}^3\epsilon_{nm}^3},
\\
\gamma_{ml}^n 
&=
\dfrac{1}{2}
\left(
\dfrac{2}{\epsilon_{nl}\epsilon_{nm}} 
-
\dfrac{\epsilon_{nl}}{\epsilon_{nm}^3}
-
\dfrac{\epsilon_{nm}}{\epsilon_{nl}^3}
\right)
=
\dfrac{(\epsilon_{nl}^2-\epsilon_{nm}^2)^2}{-2\epsilon_{nl}^3\epsilon_{nm}^3}.
\end{align}
Here, $\alpha_{ml}^n=\beta^n_{lm}$ and $\gamma^n_{ml}=\gamma^n_{lm}$ ensure $T_n^{bcd}=T_n^{bdc}$.

We remark that the first two terms of Eq.~\eqref{Tabcapp} give rise to the two-band processes of the third-order IAHE
defined by Eq.~\eqref{tau0} in the main text, as contributed by the BCQ and AQMD, respectively.
The last term of Eq.~\eqref{Tabcapp} is responsible for the three-band processes of Eq.~\eqref{tau0},
as contributed by the three-state quantum metric~\cite{MehraeenPRL} dipole (TQMD) defined by $v_n^ag^{bcd}_{nml}$.
As shown in the Lieb-lattice altermagnet and the van der Waals altermagnetic metal V$_2$Se$_2$O,
this contribution is negligibly small.

\begin{figure}[htb!]
\includegraphics[width=0.9\columnwidth]{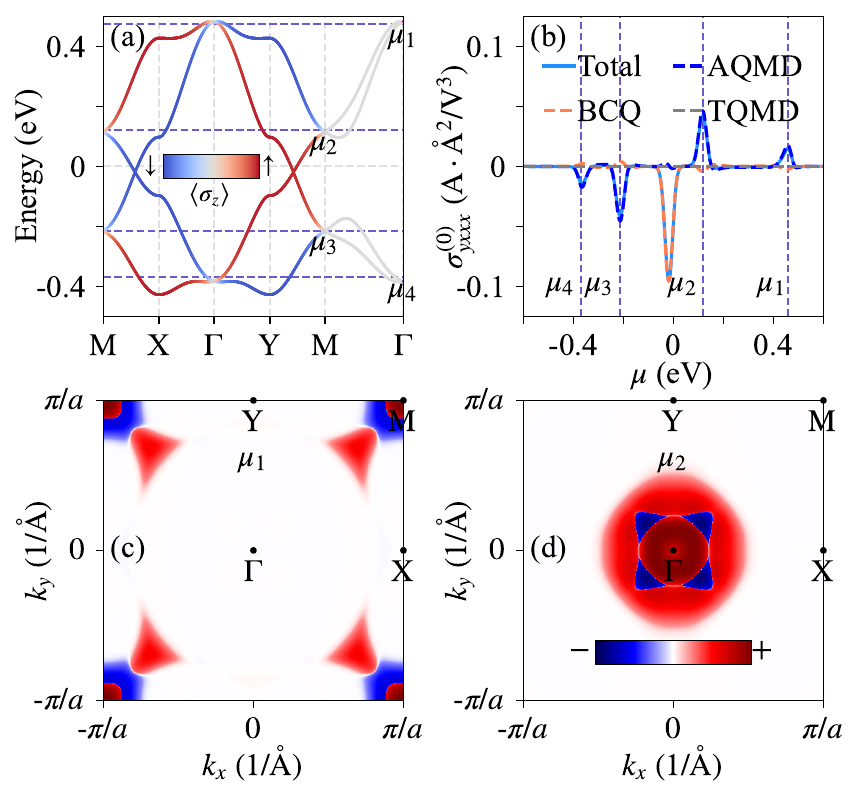}
\caption{
Third-order IAHE in 2D Lieb-lattice altermagnets with a spin-flip SOC.
(a) The band dispersions of Eq.~(\ref{LiebH}) with an additional spin-flip SOC ($\lambda_{\parallel}=0.2t_1$).
(b) The quantum-geometric decompositions of $\sigma^{(0)}_{yxxx}$.
(c-d) The normalized $\vect{k}$-resolved integrand (contributed by AQMD) for $\sigma_{yxxx}^{(0)}$ for the $\mu_1$ and $\mu_2$ peaks, respectively.
}
\label{FIG4}
\end{figure}

\subsection{Third-order IAHE from AQMD}
Here, we turn to the contribution of the AQMD to the third-order IAHE,
which becomes prominent upon introducing the spin-flip SOC.
As shown in Fig.~\ref{FIG4}(a), the inclusion of a finite spin-flip term lifts the nodal line along $\Gamma-\mathrm{M}$
and opens gaps at the previously symmetry-protected crossings,
thereby qualitatively reshaping the quantum geometry at high-symmetry momenta,
as will be clear below. Correspondingly, the conductivity $\sigma^{(0)}_{yxxx}$ for the third-order IAHE
develops multiple resonant peaks [Fig.~\ref{FIG4}(b)],
in sharp contrast to the BCQ-dominated single-peak structure in the spin-conserving case.

Using the quantum-geometric decomposition of third-order IAHE,
we find that these additional resonances are overwhelmingly governed by the AQMD,
while the BCQ part becomes subleading in this regime.
The normalized $\vect{k}$-resolved integrand further reveals that the AQMD-induced peaks are strongly localized
around high-symmetry momenta [Figs.~\ref{FIG4}(c) and \ref{FIG4}(d)],
reflecting the fact that the spin-flip SOC not only gaps the nodal line
but also generates large interband acceleration matrix elements $w^{ab}_{nm}$ around the high-symmetry momenta.
This behavior is distinct from the BCQ-dominated response,
which is primarily concentrated around generic momenta associated with altermagnetic band crossings.
In addition, the AQMD-dominated enhancement requires a large SOC
and hence the BCQ-dominated enhancement is more favorable for the experimental probing of altermagnets.

\bigskip
\subsection{Time-reversal-odd third-order EAHE}
Combining the semiclassical theory with the Boltzmann equation,
the $\mathcal{T}$-odd third-order EAHE can be similarly derived from
$j_a=-e^2/\hbar \sum_n\int_k f_n^{(2)}(\vect{E}\times\vect{\Omega}_n)_a$,
where $f_n^{(2)} = \tau^2 e^2/\hbar^2 \partial^2_{cd} f_n E_c E_d$ is the nonequilibrium Fermi distribution~\cite{KTLawPRB}
and $\vect{\Omega}_n$ is the (global) BC~\cite{Xiao2010} for the $n$th Bloch band.
By defining $j_a=\sigma_{abcd}^{(2)}E_bE_cE_d$,
the conductivity tensor for the $\mathcal{T}$-odd third-order EAHE is found to be~\cite{KTLawPRB}
\begin{align}
\sigma_{abcd}^{(2)} = \tau^2 \dfrac{e^4}{\hbar^3} \sum_{nm} \int_k f_n \partial^2_{cd} \Omega^{ab}_{nm},
\label{tau2}
\end{align}
which shows a quantum geometric origin of BCQ~\cite{KTLawPRB, BCQ2, BCQ3, BCQ4} in the form of $\partial^2_{cd} \Omega^{ab}_{nm}$.


\begin{thebibliography}{00}
\bibitem{footnoteTeven}
Note that the $\mathcal{T}$-even third-order EAHE,
arising from the quantum metric quadrupole~\cite{BCQ4, QMQ1, QMQ2},
can be distinguished from the intrinsic one in experiments and will not be discussed in this work.
\bibitem{footnote3}
By dimensional analysis, we find $\sigma^{(0)}_{abcd}/\sigma^{(2)}_{abcd}\sim\hbar^2/(\tau\mu)^2$,
where $\mu$ is the chemical potential controlled by the doping.
In a moderately dirty altermagnetic metal with~\cite{alter2024, alter2026} $\tau \sim 10^{-14} \mathrm{s}$,
we find that $\sigma^{(0)}_{abcd}/\sigma^{(2)}_{abcd} \sim 40$ with a light doping $\mu \sim 10^{-2} \mathrm{eV}$.
\bibitem{spinless}
Note that both $\sigma_{abcd}^{(0)}$ and $\kappa_{abc}$ are spinless.
\bibitem{footnote1}
Without the assumption of $\vect{N} \perp \vect{j}$ and $\vect{N} \perp \vect{E}$,
the spin point group $^26/^2m$ can support the elements $\sigma^{(0)}_{yzxx}$ and $\sigma^{(0)}_{yzxy}$ while the spin point group
$^2/^2m^2m^1m$ can support the elements $\sigma^{(0)}_{yzxx}=-\sigma^{(0)}_{zyxy}=-\sigma^{(0)}_{zxxy}$ in presence of SOC.
\end{thebibliography}
\end{document}